\documentclass[journal,twoside]{IEEEtran}

\usepackage{color}
\usepackage{bbm}
\usepackage{url}
\usepackage[cmex10]{amsmath}
\usepackage{amsfonts,amssymb}
\usepackage{dsfont}
\usepackage{graphicx}
\usepackage[hang]{subfigure}
\usepackage{citesort}
\usepackage{mathtools}
\usepackage{bbm}
\usepackage{fixmath}
\usepackage{tcolorbox}
\usepackage{balance}

\usepackage{ntheorem}
\theoremstyle{plain}
\theoremheaderfont{\itshape}\theorembodyfont{\itshape}
\theoremseparator{.}
\newtheorem{theorem}{Theorem}
\newtheorem{corollary}{Corollary}
\newtheorem{lemma}{Lemma}
\theoremheaderfont{\itshape}\theorembodyfont{\upshape}
\theoremseparator{.}
\newtheorem{definition}{Definition}
\newtheorem{remark}{Remark}

\newcommand{\N}{\mathbb{N}}
\newcommand{\Q}{\mathbb{Q}}
\newcommand{\R}{\mathbb{R}}

\newcommand{\sA}{\mathcal{A}}
\newcommand{\sC}{\mathcal{C}}
\newcommand{\sE}{\mathcal{E}}
\newcommand{\sM}{\mathcal{M}}
\newcommand{\sO}{\mathcal{O}}
\newcommand{\sP}{\mathcal{P}}
\newcommand{\sX}{\mathcal{X}}
\newcommand{\sY}{\mathcal{Y}}
\newcommand{\CH}{\mathcal{CH}}
\newcommand{\fT}{\mathfrak{T}}
\newcommand{\Mopt}{\sM_{\text{opt}}(\sX;\sY)}
\newcommand{\CHcomp}{\mathcal{CH}_c(\sX;\sY)}
\newcommand{\Pcomp}{\mathcal{P}_c(\sX)}
\newcommand{\Popt}{\sP_{\text{opt}}}

\newcommand{\addspace}{\vspace*{0.25\baselineskip}}

\begin{document}

\title{Algorithmic Computability and Approximability of Capacity-Achieving Input Distributions}
\author{Holger Boche,~\IEEEmembership{Fellow,~IEEE}, 
	Rafael F. Schaefer,~\IEEEmembership{Senior Member,~IEEE}, and
	H. Vincent Poor,~\IEEEmembership{Life Fellow,~IEEE} 
	\thanks{This work of H. Boche was supported in part by the German Federal Ministry of Education and Research (BMBF) within the national initiative on 6G Communication Systems through the research hub \emph{6G-life} under Grant 16KISK002, within the national initiative on \emph{Post Shannon Communication (NewCom)} under Grant 16KIS1003K, and the project \emph{Hardware Platforms and Computing Models for Neuromorphic Computing (NeuroCM)} under Grant 16ME0442. It has further received funding by the Bavarian Ministry of Economic Affairs, Regional Development and Energy as part of the project \emph{6G Future Lab Bavaria}. This work of R. F. Schaefer was supported in part by the BMBF within NewCom under Grant 16KIS1004 and 6G-life under Grant 16KISK001K as well as in part by the German Research Foundation (DFG) under Grant SCHA 1944/6-1. This work of H. V. Poor was supported by the U.S. National Science Foundation under Grant CCF-1908308. This article was presented in part at the IEEE International Symposium on Information Theory (ISIT), Espoo, Finland, June 2022~\cite{BocheSchaeferPoor-2022-ISIT-BlahutArimoto}.
	}
	\thanks{Holger Boche is with the Institute of Theoretical Information Technology, Technical University of Munich, the BMBF Research Hub 6G-life, the Munich Quantum Valley (MQV), 80290 Munich, Germany, and the Excellence Cluster Cyber Security in the Age of Large-Scale Adversaries (CASA), Ruhr University Bochum, 44801 Bochum, Germany (email: boche@tum.de).}
	\thanks{R. F. Schaefer is with the Chair of Information Theory and Machine Learning, Technische Universit\"at Dresden, the BMBF Research Hub 6G-life, the Cluster of Excellence ``Centre for Tactile Internet with Human-in-the-Loop (CeTI)'', and the 5G Lab Germany, Technical University of Dresden, 01069 Dresden, Germany   (E-mail: rafael.schaefer@tu-dresden.de).}
	\thanks{H. Vincent Poor is with the Department of Electrical and Computer Engineering, Princeton University, Princeton, NJ 08544, USA (email: poor@princeton.edu).}         
}

\IEEEoverridecommandlockouts
\maketitle

% ========================================================================================
% ========================================================================================
% ========================================================================================
\begin{abstract}
	The capacity of a channel can usually be characterized as a maximization of certain entropic quantities. From a practical point of view it is of primary interest to not only compute the capacity value, but also to find the corresponding optimizer, i.e., the capacity-achieving input distribution. This paper addresses the general question of whether or not it is possible to find algorithms that can compute the optimal input distribution depending on the channel. For this purpose, the concept of Turing machines is used which provides the fundamental performance limits of digital computers and therewith fully specifies which tasks are algorithmically feasible in principle. It is shown for discrete memoryless channels that it is impossible to algorithmically compute the capacity-achieving input distribution, where the channel is given as an input to the algorithm (or Turing machine). Finally, it is further shown that it is even impossible to algorithmically approximate these input distributions.
\end{abstract}

\begin{IEEEkeywords}
	Capacity-achieving input distribution, Turing machine, computability, approximability.
\end{IEEEkeywords}

% ======================================================================================================
% ======================================================================================================
% ======================================================================================================
\section{Introduction}
\label{sec:intro}

The capacity of a channel describes the maximum rate at which a sender can reliably transmit a message over a noisy channel to a receiver. Accordingly, the capacity is a function of the channel and is usually expressed by entropic quantities that are maximized over all possible input distributions. To this end, a (numerical) evaluation of the capacity and a characterization of the optimal input distribution that maximizes the capacity expression are important and common tasks in information and communication theory. To date, for discrete memoryless channels (DMCs) no general closed form solutions for the capacity expressions or the corresponding optimal input distribution as a function of the channel are known. Therefore, several approaches have been proposed to algorithmically compute the capacity and also (implicitly) the corresponding optimizer. Such a numerical simulation and computation on digital computers has been already proposed by Shannon, Gallager, and Berlekamp in \cite{ShannonGallagerBerlekamp-1967-BoundsErrorProbability} and Blahut in \cite{Blahut-1972-Computation}. This is an interesting and challenging task for digital computers which can be seen already for the binary symmetric channel with rational crossover probability $p$ whose capacity is a transcendental number\footnote{An \emph{algebraic number} is a number that is a root of a non-zero polynomial with integer coefficients. A \emph{transcendental number} is a number that is not algebraic, i.e., it is not a root of any non-zero integer polynomial.} in general except for the trivial case $p=\frac{1}{2}$ (see also the appendix for a detailed discussion on this). Thus, an exact computation of the capacity value is not possible on a digital computer as any practical algorithm must stop after a finite number of computation steps and, therefore, only an approximation of the capacity value is possible. From a practical point of view, this is not a problem since there are algorithms that take the rational crossover probability $p$ and a given approximation error $\frac{1}{2^n}$ with $n\in\N$ as inputs and stop when a rational number is calculated whose approximation error to the corresponding capacity is smaller than the required approximation error $\frac{1}{2^n}$.

A famous iterative algorithm for the computation of the capacity of an arbitrary DMC was independently proposed in 1972 by Blahut \cite{Blahut-1972-Computation} and Arimoto \cite{Arimoto-1972-Algorithm}, where the former further presented a corresponding algorithm for the computation of the rate-distortion function. This iterative algorithm is now referred to as the \emph{Blahut-Arimoto algorithm}. It was further studied by Csisz\'ar  \cite{Csiszar-1974-ComputationRateDistortion} and later generalized by Csisz\'ar and Tusn\'ady \cite{CsiszarTusnady-1984-InformationGeometryAlternatingMinimization}. The Blahut-Arimoto algorithm also appears in introductory textbooks on information theory such as \cite{CoverThomas06ElementsInformationTheory} and \cite{Yeung08InformationTheoryNetworkCoding}. Since then, the Blahut-Arimoto algorithm has been extensively studied and extended to various scenarios, cf. for example \cite{Dupuis-2004-BlahutArimoto,Vontobel-2008-BlahutArimotoFSC,OechteringAnderssonSkoglund-2009-ArimotoBlahutBidirectional,NaissPermuter-2013-BlahutArimotoDirected,Trillingsgaard-2013-BlahutArimoto,Ugut-2017-GeneralizationBlahutArimoto,LiCai-2019-BlahutArimotoClassicalQuantum,Ramakrishnan-2020-QuantumBlahutArimoto}. It further has served as the basis for the computation of the optimal input distribution in various patents such as \cite{ungerboeck2011Huffmanshaping} and \cite{zeitler2013relay}.

Blahut motivated his studies in \cite{Blahut-1972-Computation} by the desire to use digital computers, which were becoming more and more powerful at this time, for the numerical computation of the capacity of DMCs. Since the seminal works \cite{Blahut-1972-Computation} and \cite{Arimoto-1972-Algorithm}, digital computers have been extensively used in information and communication theory to simulate and evaluate the performance of communication systems. Not surprisingly, higher-layer network simulations on high performance computers has become a commonly used approach for the design of practical systems. A critical discussion on this trend is given in \cite{EphremidesHajek-1998-InformationTheoryCommunicationNetworks}.

In this paper, we address the issue of computing the optimal input distribution from a fundamental algorithmic point of view by using the concept of a \emph{Turing machine} \cite{Turing-1936-ComputableNumbersEntscheidungsproblem,Turing-1937-ComputableNumbersEntscheidungsproblemCorrection,Weihrauch-2000-ComputableAnalysis} and the corresponding \emph{computability framework}. The Turing machine is a mathematical model of an abstract machine that manipulates symbols on a strip of tape according to certain given rules. It can simulate any given algorithm and therewith provides a simple but very powerful model of computation. Turing machines have no limitations on computational complexity, unlimited computing capacity and storage, and execute programs completely error-free. They are further equivalent to the von Neumann-architecture without hardware limitations and the theory of recursive functions, cf. also \cite{AvigadBrattka-2014-ComputabilityAnalysis,Godel-1930-VollstandigkeitAxiome,Godel-1934-UndecidablePropositions,Kleene-1952-IntroductionMetamathematics,Minsky-1961-RecursiveUnsolvability}. Accordingly, Turing machines provide fundamental performance limits for today's digital computers and are the ideal concept to study whether or not such computation tasks can be done algorithmically in principle.

Communication from a computability or algorithmic point of view has attracted some attention recently. In \cite{BocheSchaeferPoor-2019-TIFS-ComputabilitySecureCommunicationIdentification} the computability of the capacity functions of the wiretap channel under channel uncertainty and adversarial attacks is studied. The computability of the capacity of finite state channels is studied in \cite{BocheSchaeferPoor-2020-CIS-NonComputabilityFSC} and of non-i.i.d. channels in \cite{BocheSchaeferPoor-2019-ITW-NonIID}. These works have in common that they study capacity functions of various communication scenarios and analyze the algorithmic computability of the capacity function itself. While for DMCs the capacity function is a computable continuous function and therewith indeed algorithmically computable \cite{PourElRichards-2017-ComputabilityAnalysisPhysics,BocheSchaeferBaurPoor-2019-TSP-ComputabilitySKGAuthentication}, this is no longer the case for certain multi-user scenarios or channels with memory. However, they do not consider the computation of the optimal input distributions which, to the best of our knowledge, has not been studied so far from a fundamental algorithmic point of view. In addition, even if the capacity is computable, it is still not clear whether or not the corresponding optimal input distributions can be algorithmically computed.

We consider finite input and output alphabets. Due to the properties of the mutual information, the set of capacity-achieving input distributions is mathematically well defined for every DMC and so are all functions that map every channel to a corresponding capacity-achieving input distribution. A practically relevant question is now whether or not these functions are also algorithmically well defined. With this we mean whether or not it is possible to find at least one function that can be implemented by an algorithm (or Turing machine). This is equivalent  to the question of whether or not a Turing machine exists that takes a computable channel as input and subsequently computes an optimal input distribution of this channel.

In this paper, we give a negative answer to the question above by showing that it is in general impossible to find an algorithm (or Turing machine) that is able to compute the optimal input distribution when the channel is given as an input. To this end, we first introduce the computability framework based on Turing machines in Section \ref{sec:computability}. The communication system model and the Blahut-Arimoto algorithm are subsequently introduced in Section \ref{sec:system}. In Section \ref{sec:system_problem} we study the computability of an optimal input distribution and show that all functions that map channels to their corresponding optimal input distributions are not Banach-Mazur computable and therewith also not Turing computable. As a consequence, there is no algorithm (or Turing machine) that is able to compute the optimizer, i.e., the capacity-achieving input distribution. Subsequently, it is shown in Section \ref{sec:system_approx} that it is further not even possible to algorithmically approximate the optimizer, i.e., the capacity-achieving input distribution, within a given tolerated error. Finally, a conclusion is given in Section \ref{sec:conclusion}.

% ======================================================================================================
\subsection*{Notation}

Discrete random variables are denoted by capital letters and their realizations and ranges by lower case and calligraphic letters, respectively; all logarithms and information quantities are taken to the base 2; $\N$, $\Q$, and $\R$ are the sets of non-negative integers, rational numbers, and real numbers; $\sP(\sX)$ denotes the set of all probability distributions on $\sX$ and $\CH(\sX;\sY)$ denotes the set of all stochastic matrices (channels) $\sX\rightarrow\sP(\sY)$; the binary entropy is denoted by $h_2(p)=-p\log p - (1-p)\log(1-p)$ and $I(X;Y)$ denotes the mutual information between the input $X$ and the output $Y$ which we interchangeably also write as $I(p,W)$ to emphasize the dependency on the input distribution $p\in\sP(\sX)$ and the channel $W\in\CH(\sX;\sY)$; the $\ell_1$-norm is denoted by $\|\cdot\|_{\ell_1}$.

% ========================================================================================
% ========================================================================================
% ========================================================================================
\section{Computability Framework}
\label{sec:computability}

We first introduce the computability framework based on Turing machines which provides the needed background. Turing machines are extremely powerful compared to state-of-the-art digital signal processing (DSP) and field gate programmable array (FPGA) platforms and even current supercomputers. It is the most general computing model and is even capable of performing arbitrary exhaustive search tasks on arbitrary large but finite structures. The complexity can even grow faster than double-exponentially with the set of parameters of the underlying communication system (such as time, frequencies, transmit power, modulation scheme, number of  antennas, etc.). 

In what follows, we need some basic definitions and concepts of computability which are briefly reviewed. The concept of computability and computable real numbers was first introduced by Turing in \cite{Turing-1936-ComputableNumbersEntscheidungsproblem} and \cite{Turing-1937-ComputableNumbersEntscheidungsproblemCorrection}. 

Recursive functions $f:\N\rightarrow\N$ map natural numbers into natural numbers and are exactly those functions that are computable by a Turing machine. They are the smallest class of partial functions that includes the primitive functions (i.e., the constant function, successor function, and projection function) and is further closed under composition, primitive recursion, and minimization. For a detailed introduction, we refer the reader to \cite{PourElRichards-2017-ComputabilityAnalysisPhysics} and \cite{Soare-1987-RecursivelyEnumerableSetsDegrees}. With this, we call a sequence of rational numbers $(r_n)_{n\in\N}$ 
a \emph{computable sequence} if there exist recursive functions $a,b,s:\N\rightarrow\N$ with $b(n)\neq0$ for all $n\in\N$ and
\begin{equation}
	\label{eq:computability_comp1}
	r_n= (-1)^{s(n)}\frac{a(n)}{b(n)}, \qquad n\in\N;
\end{equation}
cf.  \cite[Def. 2.1 and 2.2]{Soare-1987-RecursivelyEnumerableSetsDegrees} for a detailed treatment. A real number $x$ is said to be computable if there exists a computable sequence of rational numbers $(r_n)_{n\in\N}$ and a recursive function $\varphi$ such that we have for all $M\in\N$
\begin{equation}
	\label{eq:computability_comp2}
	|x-r_n|<2^{-M}
\end{equation}
for all $n\geq \varphi(M)$. Thus, the computable real $x$ is represented by the pair $((r_n)_{n\in\N},\varphi)$. Note that a computable real number usually has multiple different representations. For example, there are multiple algorithms known for the computation of $\frac{1}{\pi}$ or $e^{-1}$. This form of convergence \eqref{eq:computability_comp2} with a computable control of the approximation error is called \emph{effective convergence}. 

For the definition of a \emph{computable sequence of computable real numbers} we need the following definition as in \cite{PourElRichards-2017-ComputabilityAnalysisPhysics}. 

\begin{definition}
	Let $(x_{nk})_{n,k\in\N}$ be a double sequence of real numbers and $(x_n)_{n\in\N}$ a sequence of real numbers such that $x_{nk}\rightarrow x_n$ for each $n$ as $k\rightarrow\infty$.  We say that $x_{nk}\rightarrow x_n$ effectively in $k$ and $n$ if there is a recursive function $\varphi:\N\times\N\rightarrow\N$ such that for all $n,N$ we have $k \geq \varphi(n,N)$ implies
	\begin{equation*}
		|x_{nk}-x_n|\leq 2^{-N}.
	\end{equation*}
\end{definition}

With this, we get the following definition.

\begin{definition}
	A sequence of computable real numbers $(x_n)_{n\in\N}$ is a \emph{computable sequence} if there is a computable double sequence of rational numbers $r_{nk}$ such that $r_{nk}\rightarrow x_n$ as $k\rightarrow\infty$, effectively in $k$ and $n$. 
\end{definition}

This can alternatively be stated as follows, cf. also \cite{PourElRichards-2017-ComputabilityAnalysisPhysics}. A sequence of computable real numbers $(x_n)_{n\in\N}$ is a computable sequence if there is a computable double sequence of rational numbers $(r_{nk})_{n,k\in\N}$ such that 
\begin{equation*}
	|r_{nk}-x_n|\leq 2^{-k}, \qquad \text{for all }k \text{ and } n.
\end{equation*}

Note that if a computable sequence of computable real numbers $(r_n)_{n\in\N}$ converges effectively to a limit $x$, then $x$ is a computable real number, cf. \cite{PourElRichards-2017-ComputabilityAnalysisPhysics}. Furthermore, the set $\R_c$ of all computable real numbers is closed under addition, subtraction, multiplication, and division (excluding division by zero). We denote the set of computable real numbers by $\R_c$. Based on this, we define the set of computable probability distributions $\sP_c(\sX)$ as the set of all probability distributions $P_X\in\sP(\sX)$ such that $P_X(x)\in\R_c$ for all $x\in\sX$. Further, let $\CHcomp$ be the set of all computable channels, i.e., for a channel $W:\sX\rightarrow\sP(\sY)$ we have $W(\cdot|x)\in\sP_c(\sY)$ for every $x\in\sX$. 

\begin{definition}
	\label{def:borel}
	A function $f:\R_c\rightarrow\R_c$ is called \emph{Borel-Turing computable} if there exists an algorithm or Turing machine $\fT_f$ such that $\fT_f$ obtains for every $x$ an arbitrary representation $((r_n)_{n\in\N},\varphi)$ for it as input and then computes a representation $((\hat{r}_n)_{n\in\N},\hat{\varphi})$ for $f(x)$.
\end{definition}

\begin{remark}
	\label{rem:turing1}
	Borel-Turing computability characterizes exactly the behavior that is expected when functions are simulated and evaluated on digital hardware platforms. A program for the computation of $f(x)$ must receive  a representation $((r_n)_{n\in\N},\varphi)$ for the input $x$. Based on this, the program computes the representation $((\hat{r}_n)_{n\in\N},\hat{\varphi})$ for $f(x)$. This means that if $f(x)$ needs to be computed with a tolerated approximation error of $\frac{1}{2^M}$, then it is sufficient to compute the rational number $\hat{r}_{\hat{\varphi}(M)}$ and the corresponding Turing machine outputs $\hat{r}_{\hat{\varphi}(M)}$. For example, this is done and further discussed for the function $f(x)=e^{-x}$, $x\in[0,1]$, $x\in\R_c$ in Appendix \ref{app:function}. 
\end{remark}

\begin{remark}
	\label{rem:turing2}
	A practical digital hardware platform and also a Turing machine must stop after finitely many computation steps when computing a value of a function. Thus, the computed value of the function must be a rational number. As a consequence, a Turing machine can only compute rational numbers \emph{exactly}. However, it is important to note that in information and communication theory, the relevant information-theoretic functions are in general not exactly computable even for rational channel and system parameters. For example, already for $|\sX|=2$ and rational probability distribution $p\in\sP(\sX)$, $p\neq\begin{pmatrix}	\frac{1}{2}, \frac{1}{2}\end{pmatrix}$, the corresponding binary entropy $h_2(p)$ is a transcendental number and therewith not exactly computable. Even if this would be done symbolically with algebraic numbers, the binary entropy would not be computable.  As a consequence, already for the binary symmetric channel (BSC) with rational crossover probability $\epsilon\in(0,\frac{1}{2})\cap\Q$, the capacity $C_{\text{BSC}}(\epsilon)=1-h_2(\epsilon)$ is a transcendental number and therewith an exact computation of the capacity is not possible. A proof for this statement is given in Appendix \ref{app:transcendental} for completeness.
\end{remark}

There are also weaker forms of computability including \emph{Banach-Mazur computability}. In particular, Borel-Turing computability implies Banach-Mazur computability, but not vice versa. For an overview of the logical relations between different notions of computability we refer to \cite{AvigadBrattka-2014-ComputabilityAnalysis} and, for example, the introductory textbook \cite{Weihrauch-2000-ComputableAnalysis}.

\begin{definition}
	\label{def:banachmazur}
	A function $f:\R_c\rightarrow\R_c$ is called \emph{Banach-Mazur computable} if $f$ maps any given computable sequence $(x_n)_{n\in\N}$ of computable real numbers into a computable sequence $(f(x_n))_{n\in\N}$ of computable real numbers.
\end{definition}

We further need the concepts of a recursive set and a recursively enumerable set as, for example, defined in \cite{Soare-1987-RecursivelyEnumerableSetsDegrees}.

\begin{definition}
	\label{def:recursive}
	A set $\sA\subset\N$ is called \emph{recursive} if there exists a computable function $f$ such that $f(x)=1$ if $x\in\sA$ and $f(x)=0$ if $x\notin\sA$. 
\end{definition}

\begin{definition}
	\label{def:recursiveenumerable}
	A set $\sA\subset\N$ is \emph{recursively enumerable} if there exists a recursive function whose range is exactly $\sA$.
\end{definition}

We have the following properties which will be crucial later for proving the desired results; cf. also \cite{Soare-1987-RecursivelyEnumerableSetsDegrees} for further details.
\begin{itemize}
	\item $\sA$ is recursive is equivalent to $\sA$ is recursively enumerable and $\sA^c$ is recursively enumerable.
	\item There exist recursively enumerable sets $\sA\subset\N$ that are not recursive, i.e., $\sA^c$ is not recursively enumerable. This means there are no computable, i.e., recursive, functions $f:\N\rightarrow\sA^c$ where for each $m\in\sA^c$ there exists an $x$ with $f(x)=m$.
\end{itemize}

% ========================================================================================
% ========================================================================================
% ========================================================================================
\section{System Model and Blahut-Arimoto Algorithm}
\label{sec:system}

Here, we introduce the communication scenario of interest and discuss the Blahut-Arimoto algorithm.

% ========================================================================================
\subsection{Communication System Model}
\label{sec:system_model}

We consider a point-to-point channel with one transmitter and one receiver which defines the most basic communication scenario. Let $\sX$ and $\sY$ be finite input and output alphabets. Then the channel is given by a stochastic matrix $W:\sX\rightarrow\sP(\sY)$ which we also equivalently write as $W\in\CH(\sX;\sY)$. The corresponding DMC is then given by $W^n(y^n|x^n)\coloneqq\prod_{i=1}^nW(y_i|x_i)$ for all $x^n\in\sX^n$ and $y^n\in\sY^n$. 

\begin{definition}
	\label{def:code}
	An $(M_n,E_n,D_n)$-\emph{code} $\sC_n(W)$ of blocklength $n\in\N$ for the DMC $W\in\CH(\sX;\sY)$ consists of an encoder $E_n:\sM_n\rightarrow\sX^n$ at the transmitter with a set of messages $\sM_n\coloneqq\{1,...,M_n\}$ and a decoder $D_n:\sY^n\rightarrow\sM_n$ at the receiver.  
\end{definition}

The transmitted codeword needs to be decoded reliably at the receiver. To model this requirement, we define the \emph{average probability of error} as
\begin{equation*}
	\bar{e}_n \coloneqq \frac{1}{|\sM_n|}\sum_{m\in\sM_n}\sum_{y^n:D_n(y^n)\neq m}W^n(y^n|x^n_m)
\end{equation*}
and the \emph{maximum probability of error} as
\begin{equation*}
	e_{\text{max},n} \coloneqq \max_{m\in\sM_n}\sum_{y^n:D_n(y^n)\neq m}W^n(y^n|x^n_m)
\end{equation*}
with $x_m^n=E_n(m)$ the codeword for message $m\in\sM_n$.

\begin{definition}
	\label{def:achievable}
	A rate $R>0$ is called \emph{achievable} for the DMC $W$ if there exists a sequence $(\sC_n(W))_{n\in\N}$ of $(M_n,E_n,D_n)$-codes such that we have $\frac{1}{n}\log M_n\geq R$ and $\bar{e}_n\leq\epsilon_n$ (or $e_{\text{max},n}\leq\epsilon_n$, respectively) with $\epsilon_n\rightarrow0$ as $n\rightarrow\infty$. The \emph{capacity} $C(W)$ of the DMC $W$ is given by the supremum of all achievable rates $R$.
\end{definition}

The capacity of the DMC has been established and goes back to the seminal work of Shannon \cite{Shannon-1948-AMathematicalTheoryOfCommunication}.

\begin{theorem}
	\label{the:capacity}
	The capacity $C(W)$ of the DMC $W$ under both the average and maximum error criteria is
	\begin{equation}
		\label{eq:system_capacity}
		C(W) = \max_XI(X;Y) = \max_{p\in\sP(\sX)}I(p,W).
	\end{equation}
\end{theorem}

The capacity of a channel characterizes the maximum transmission rate at which the users can reliably communicate with vanishing probability of error. Note that for DMCs, there is no difference in the capacity whether the average error or the maximum error criterion is considered.

\begin{remark}
	\label{rem:capacity}
	Capacity expressions such as \eqref{eq:system_capacity} for the point-to-point channel have further been established for various multi-user communication scenarios, cf. for example \cite{ElGamalKim-2011-NetworkInformationTheory} and references therein. They all have in common that these are characterized by entropic quantities.
\end{remark}

% ========================================================================================
\subsection{Blahut-Arimoto Algorithm}
\label{sec:system_blahutarimoto}

The Blahut-Arimoto algorithm as initially proposed in \cite{Blahut-1972-Computation} and \cite{Arimoto-1972-Algorithm} tackles the problem of numerically computing the capacity of DMCs with finite input and output alphabets. This algorithm is an alternating optimization algorithm, which has become a standard technique of convex optimization. It has the advantage that it exploits the properties of the mutual information to obtain a simple method to compute the capacity.

For a DMC $W$, the algorithm computes the following two quantities at the $n$-th iteration:
\begin{enumerate}
	\item an input distribution $p_n=p_n(W)$
	\item an approximation to the capacity given by the mutual information $I(p_n,W)$ for this input distribution.
\end{enumerate}
This means that the algorithm computes a sequence $p_0(W)$, $I(p_0,W)$, $p_1(W)$, $I(p_1,W)$, ... , $p_n(W)$, $I(p_n,W)$, ... where each element in the sequence is a function of the previous ones except the initial input distribution $p_0(W)$ which is arbitrarily chosen. It is clear that the sequence $(p_n(W))_{n\in\N}$ of computable input distributions is a function of the initial input distribution $p_0(W)$. The same is true for the sequence $(I(p_n(W),W))_{n\in\N}$.

For the sequence $p_0(W)$, $p_1(W)$, ... it is shown in \cite{Arimoto-1972-Algorithm,Blahut-1972-Computation,Csiszar-1974-ComputationRateDistortion} that it always contains a convergent subsequence and that all these convergent subsequences converge to a corresponding optimal input distribution. First, the existence of a limit $p_*=p_*(W)\in\sP(\sX)$ of this subsequence is shown by the Bolzano–Weierstra\ss{} theorem, cf. for example \cite{Bartle-2000-RealAnalysis}. Subsequently, it is shown that this limit must be an optimal input distribution, i.e., $p_*\in\Popt(W)$ with
\begin{equation}
	\label{eq:popt}
	\Popt(W) = \big\{p\in\sP(\sX):I(p,W)=C(W)\big\}
\end{equation}
the set of optimal input distributions. The Bolzano-Weierstra\ss{} theorem is a simple technique to show the existence of solutions of certain problems, but, in general, it does not provide an algorithm to compute this solution; in this case the optimal input distribution as a function of the channel. 

For the capacity, a stopping criterion is provided, i.e., we can choose a certain approximation error $\frac{1}{2^M}>0$, $M\in\N$, and the algorithm stops if this tolerated error is satisfied so that the computed value $I(p_n,W)$ is within this error to the actual capacity $C(W)$, i.e.,
\begin{equation*}
	\big|C(W)-I(p_n,W)\big|<\frac{1}{2^M},
\end{equation*} 
see \cite[Corollary 1]{Arimoto-1972-Algorithm} for a stopping condition for iterations of the capacity estimation.

On the other hand, although it has been studied in \cite{Arimoto-1972-Algorithm,Blahut-1972-Computation,Csiszar-1974-ComputationRateDistortion}, a stopping criterion for the optimizer, i.e., the optimal input distribution, has not been given in \cite{Arimoto-1972-Algorithm,Blahut-1972-Computation,Csiszar-1974-ComputationRateDistortion}, i.e., we cannot control when the algorithm should stop for a given maximum tolerable error. Such a stopping criterion could similarly be defined, e.g., when 
\begin{equation}
	\big\|p_*-p_n(W)\big\|_{\ell_1}<\frac{1}{2^M}
	\label{eq:A}
\end{equation}
is satisfied with $p_*\in\Popt(W)$ a capacity-achieving input distribution and further a computable upper bound for the speed of convergence is given. Surprisingly, to date such a stopping criterion has not been found although it has been studied extensively and further has played a crucial role in various patents such as \cite{ungerboeck2011Huffmanshaping} and \cite{zeitler2013relay}. In particular, our results even show that such a stopping criterion cannot exist! We will come back to this issue in more detail in the following subsection. We will also show that the optimization over possible starting points, i.e., the computable choice of a suitable starting point for the iterating algorithm depending on the input channel $W$ does not yield a solution, i.e., a computable stopping criterion such as \eqref{eq:A}.

In fact, both seminal papers \cite{Blahut-1972-Computation} and \cite{Arimoto-1972-Algorithm} do not only aim at computing the capacity, but also propose an algorithm for the computation of a sequence of input distributions $p_n\in\sP(\sX)$ and study the convergence to a maximum $p_*\in\Popt(W)$ for a fixed channel $W$, i.e., 
\begin{equation}
	I(p_*,W) = C(W) = \max_{p\in\sP(\sX)}I(p,W).
	\label{eq:B}
\end{equation}
They state that a suitable subsequence $(p_{n_l})_{l\in\N}$ converges to an optimizer, but without providing a stopping criterion. That this is problematic has been realized afterwards by Csisz\'ar who explicitly states in \cite{Csiszar-1974-ComputationRateDistortion} that there is no stopping criterion for the computation of the optimizer. In particular, Arimoto considered the problem of estimating the convergence speed for the calculation of the input distribution. However, only under certain conditions on the input distribution was he able to show monotonicity \cite[Theorem 2]{Arimoto-1972-Algorithm} and properties of the rate of convergence \cite[Theorem 3]{Arimoto-1972-Algorithm}. From an algorithmic point of view, these results are not useful, since i) \cite[Theorem 2]{Arimoto-1972-Algorithm} shows only monotonicity, ii) \cite[Theorem 3]{Arimoto-1972-Algorithm} shows only the existence of certain parameters, but no explicit construction of them so that the result is non-constructive, and iii) there is no algorithm known to test these conditions, i.e., it is not verifiable whether \cite[Theorem 2]{Arimoto-1972-Algorithm} and \cite[Theorem 3]{Arimoto-1972-Algorithm} are applicable. Csisz\'ar studied in \cite{Csiszar-1974-ComputationRateDistortion} the question of understanding and calculating the convergence speed of the input distribution. However, he was only able to show convergence but not to calculate the rate of convergence. Accordingly, the corresponding proof of existence is non-constructive.

% ========================================================================================
\section{Computability of an Optimal Input Distribution}
\label{sec:system_problem}

\begin{figure}
	\centering
	\scalebox{1}{\includegraphics{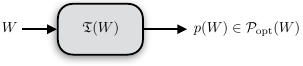}}
	\caption{Computation of the optimal input distribution. The Turing machine $\fT$ obtains a description of the channel $W$ as input and outputs a description of an optimal input distribution $p(W)$ (as an implementation of the function $G$).}
	\label{fig:1}
\end{figure}

The capacity $C(W) = \max_{p\in\sP(\sX)}I(p,W)$ of the DMC $W$, cf. \eqref{eq:system_capacity}, is given by a maximization problem, where the mutual information $I(p,W)$ is maximized over all possible input distributions $p\in\sP(\sX)$. Since $I(p,W)$ is continuous in $(p,W)$, concave in the input distribution $p$, and convex in the channel $W$, there exists for every channel $W\in\CH(\sX;\sY)$ at least one optimal input distribution $p_*(W)\in\Popt(W)$. Note that the set $\Popt(W)$ is a convex set for each channel $W$. Now, we can choose for every channel $W\in\CH(\sX;\sY)$ such a capacity-achieving input distribution $p_*=p_*(W)$. Then $F(W)=p_*(W)$ is a mathematically well defined function of the form
\begin{equation}
	\label{eq:system_f}
	F: \CH(\sX;\sY) \rightarrow \sP(\sX)
\end{equation}
which maps every channel to an optimal input distribution for this channel. We call $F$ an optimal assignment function and denote by $\Mopt$ the set of all these functions. The set $\Mopt$ is of crucial practical importance and, in particular, it would be interesting to find functions $F\in\Mopt$ that can be described algorithmically. Note that in general, this function $F$ does not need to be unique and there can be infinitely many such functions. Further, for computable channels $W\in\CHcomp$ we always have $F(W)\in\sP_c(\sX)$. 

\begin{remark}
	\label{rem:function}
	From a practical point of view it is interesting to understand whether or not there exists a function $F$ with $F(W)\in\Popt(W)$ for all $W\in\CHcomp$ that is Borel-Turing computable. Since exactly in this case there is an algorithm (or Turing machine) that takes the channel $W\in\CHcomp$ as an input and computes a corresponding capacity-achieving input distribution $F(W)=p_*(W)\in\Popt(W)$. It is clear that we consider only computable channels $W\in\CHcomp$ as inputs for the Turing machine as it can operate work only with such inputs. More specifically, for $W\in\CHcomp$ such a Turing machine takes an arbitrary representation of $W$ as input, i.e., $W(y|x)$ is given by a representation $((r_n(x,y))_{n\in\N},\varphi^{(x,y)})$ for all $x\in\sX$, $y\in\sY$. This means that for all $x\in\sX$, $y\in\sY$ we have for all $N\in\N$
	\begin{equation*}
		\big|W(y|x) - r_n(x,y)\big| < \frac{1}{2^N}
	\end{equation*}
	for all $n\geq\varphi^{(x,y)}(N)$. As a result, the Turing machine computes a representation of $F(W)\in\Popt(W)$, i.e., $((r_n^*(x))_{n\in\N},\varphi^{(*,x)})$ is a representation of $p_*(x)$, $x\in\sX$, with $F(W)=p_*=\begin{pmatrix} p_*(1), \dots ,p_*(|\sX|)\end{pmatrix}$. Thus, for all $x\in\sX$ it holds that for all $N\in\N$ 
	\begin{equation}
		\label{eq:NrA}
		\big|p_*(x) - r_n^*(x)\big| < \frac{1}{2^N}
	\end{equation}
	for all $n\geq\varphi^{(*,x)}(N)$. 
\end{remark}

Accordingly, in the following we will address this question in detail and study whether or not it is possible to find such  a Turing machine that computes a capacity-achieving input distribution for a given channel. 
\addspace

\begin{tcolorbox}[colback=white,boxrule=0.125ex]
	{\bf Question 1:} 
	Let $\sX$ and $\sY$ be finite input and output alphabets. Is there an algorithm (or Turing machine) $\fT$ that takes an arbitrary representation of $W\in\CHcomp$ as an input and computes a description of $p_*(W)\in\Popt(W)$?
\end{tcolorbox}

\begin{remark}
	\label{rem:question1}
	Question 1 is visualized in Fig.~\ref{fig:1} and formalizes exactly what we would require from an algorithmic construction of optimal input distributions on digital hardware platforms. From a practical point of view, a simulation on digital hardware must stop after a finite number of computations. Usually, it should stop if for $W\in\CHcomp$ the computed approximation of an input distribution $p_*(W)\in\Popt(W)$ satisfies a given but fixed approximation error. This constraint on the approximation error is exactly modeled by the representation of $p_*(W)$. If the representation $((r_n^*(x))_{n\in\N},\varphi^{(*,x)})$, $x\in\sX$, of $p_*(W)$ has been computed for a tolerated error $\frac{1}{2^N}$ and $r$ being the smallest natural number such that $2^r>|\sX|$, then the approximation process can be stopped after $N^*=\max_{x\in\sX}\varphi^{(*,x)}(N+r)$ steps, since we have
	\begin{align*}
		\sum_{x\in\sX}\big|p_*(x)-r_{N^*}^*(x)\big| < \sum_{x\in\sX}\frac{1}{2^{N+r}} = \frac{|\sX|}{2^{N+r}} < \frac{1}{2^N}.
	\end{align*}
	This would provide us a stopping criterion as discussed in Section~\ref{sec:system_blahutarimoto} for the Blahut-Arimoto algorithm.
\end{remark}

Now we can state the following result which provides a negative answer to Question 1 above.

\begin{theorem}
	\label{the:banachmazur}
	Let $\sX$ and $\sY$ be arbitrary but finite alphabets with $|\sX|\geq3$ and $|\sY|\geq2$. Then there is no function $F\in\Mopt$ that is Banach-Mazur computable. 
\end{theorem}
\begin{IEEEproof}
	The proof is given below in Section \ref{sec:main_comp}.
\end{IEEEproof}
\addspace

From this, we can immediately conclude the following.

\begin{corollary}
	\label{cor:banachmazur}
	There is no Turing machine $\fT$ that takes a channel $W\in\CHcomp$ as an input and computes an optimal input distribution $p\in\Popt(W)$ for this channel. 
\end{corollary}
\begin{IEEEproof}
	If such a Turing machine would exist, then the corresponding function $F$ would be Banach-Mazur computable. This is a contradiction to Theorem \ref{the:banachmazur} so that such a Turing machine cannot exist. 
\end{IEEEproof}

% ======================================================================================================
\subsection{Preliminary Considerations}
\label{sec:comp_preliminary}

Before we present the proof of Theorem \ref{the:banachmazur}, we first need to define and discuss specific channels and their optimal input distributions.

Let $\sX$ and $\sY$ be arbitrary but finite alphabets with $|\sX|=3$ and $|\sY|=2$. We define the channel
\begin{equation}
	\label{eq:comp_wstar}
	W_* = \begin{pmatrix}
		1 & 0 & 0 \\
		0 & 1 & 1 
	\end{pmatrix}
\end{equation}
and further consider the channels
\begin{equation*}
	W_{1,\mu} = \begin{pmatrix}
		1 & 0 & \mu \\
		0 & 1 & 1-\mu 
	\end{pmatrix}\quad\text{and}\quad
	W_{2,\mu} = \begin{pmatrix}
		1 & \mu & 0 \\
		0 & 1-\mu & 1 
	\end{pmatrix}
\end{equation*}
for $\mu\in(0,1)$. We define the distance between two channels $W_1,W_2\in\CH(\sX;\sY)$ based on the total variation distance as 
\begin{equation*}
	D(W_1,W_2) \coloneqq \max_{x\in\sX}\sum_{y\in\sY}\big|W_1(y|x)-W_2(y|x)\big|
\end{equation*}
and observe that
\begin{equation*}
	\lim_{\mu\rightarrow0}D(W_*,W_{1,\mu}) = \lim_{\mu\rightarrow0}D(W_*,W_{2,\mu}) = 0.
\end{equation*}
We consider the set
\begin{equation*}
	\sP_1 = \Big\{ p=(p_1,p_2,p_3)\in\sP(\sX): p_1=\frac{1}{2} \text{ and } p_2+p_3 = \frac{1}{2} \Big\}.
\end{equation*}
Then we have
\begin{equation*}
	\max_{p\in\sP(\sX)}I(p,W_*) = 1 = I(p_*,W_*)
\end{equation*}
with $p_*\in\sP_1$ arbitrary. This means $\sP_1$ is the set of all maximizing input distributions for the channel $W_*$, since
\begin{align*}
	I(p,W_*) &= p_1\cdot1\cdot\log\frac{1\cdot p_1}{p_1\cdot p_1} + p_2\cdot1\cdot\log\frac{1\cdot p_2}{p_2(p_2+p_3)} \\
	&\qquad + p_3\cdot1\cdot\log\frac{1\cdot p_3}{p_3(p_2+p_3)} \\
	&= p_1\log\frac{1}{p_1} + (p_2+p_3)\log\frac{1}{p_2+p_3} \\
	&= p_1\log\frac{1}{p_1} + (1-p_1)\log\frac{1}{1-p_1} \\
	&=h_2(p_1)
\end{align*}
where $h_2(\cdot)$ is the binary entropy function. This means that for all ${p}$ with ${p}_1\in[0,1]\backslash\{\frac{1}{2}\}$ we always have
\begin{equation*}
	I({p},W_*)<1=h_2(p_*)=I(p_*,W_*)
\end{equation*}
with $p_*\in\sP_1$ arbitrary as defined above.

Next, we define the channel
\begin{equation*}
	\hat{W} = \begin{pmatrix}
		1 & 0 & 1 \\
		0 & 1 & 0 
	\end{pmatrix}
\end{equation*}
and for $\mu\in[0,1]$ we have
\begin{equation*}
	W_{1,\mu} = (1-\mu)W_* + \mu\hat{W}.
\end{equation*}
Then for $p\in\sP(\sX)$ arbitrary, we always have
\begin{equation*}
	I(p,W_{1,\mu}) \leq (1-\mu)I(p,W_*) + \mu I(p,\hat{W}).
\end{equation*}
We now consider the set
\begin{equation*}
	\sP_2 = \Big\{ p=(p_1,p_2,p_3)\in\sP(\sX): p_2=\frac{1}{2} \text{ and } p_1+p_3 = \frac{1}{2} \Big\}.
\end{equation*}
Similarly, we can show for the channel $\hat{W}$ that
\begin{equation*}
	\max_{p\in\sP(\sX)}I(p,\hat{W}) = 1 = I(\hat{p},\hat{W})
\end{equation*}
with $\hat{p}\in\sP_2$ arbitrary. Further, we have
\begin{equation*}
	\sP_1\cap\sP_2 = \begin{pmatrix}
		\frac{1}{2} \\ \frac{1}{2} \\ 0
	\end{pmatrix}.
\end{equation*}
For $p\in\sP(\sX)$, $p\neq(\frac{1}{2},\frac{1}{2},0)$, we must have
\begin{equation*}
	I(p,W_*) < 1 \quad\text{or}\quad I(p,\hat{W})<1.
\end{equation*}
Thus, for arbitrary $p\in\sP(\sX)$ with $p\in\sP(\sX)$, $p\neq(\frac{1}{2},\frac{1}{2},0)$ we always have
\begin{align*}
	I(p,W_{1,\mu}) &\leq (1-\mu)I(p,W_*) + \mu I(p,\hat{W}) \\
	&< (1-\mu) + \mu \\
	&= 1.
\end{align*}
For
\begin{equation*}
	p_*^{(1)} = \begin{pmatrix}
		\frac{1}{2} \\ \frac{1}{2} \\ 0
	\end{pmatrix}
\end{equation*}
we have
\begin{equation*}
	I(p_*^{(1)},W_{1,\mu}) = 1
\end{equation*}
for $\mu\in[0,1]$. Consequently, for channel $W_{1,\mu}$ for $\mu\in(0,1)$ there is exactly one optimal input distribution, i.e., $\Popt(W_{1,\mu})=\{p_*^{(1)}\}$.

Similarly, one can show that for channel $W_{2,\mu}$ for $\mu\in(0,1)$ there is exactly one optimal input distribution, i.e., $\Popt(W_{2,\mu})=\{p_*^{(2)}\}$ given by
\begin{equation*}
	p_*^{(2)} = \begin{pmatrix}
		\frac{1}{2} \\ 0 \\ \frac{1}{2} 
	\end{pmatrix}.
\end{equation*}

% ======================================================================================================
\subsection{Non-Computability of an Optimal Input Distribution}
\label{sec:main_comp}

Now we are in the position to prove Theorem \ref{the:banachmazur}.
\addspace

\begin{IEEEproof}[Proof of Theorem \ref{the:banachmazur}]
We start with the case $|\sX|=3$ and $|\sY|=2$ and prove the desired result by contradiction. For this purpose, we assume that there exists a function $F\in\Mopt$ that is Banach-Mazur computable. This means that every computable sequence $(W_n)_{n\in\N}$ of computable channels $W_n\in\CHcomp$ is mapped to a computable sequence $(p_n)_{n\in\N}$ of computable input distributions $p_n\in\Pcomp$ for all $n\in\N$. For the set of optimal input distributions \eqref{eq:popt} we always have $\Popt(W)\neq\emptyset$. Further, let $F$ be an arbitrary function as in \eqref{eq:system_f} and
\begin{equation*}
	F(W) \in \Popt(W),
\end{equation*}
i.e., $F$ maps every channel to an optimal input distribution for this channel. 

For our previously defined channel $W_*$, cf. \eqref{eq:comp_wstar}, we therefore have
\begin{equation*}
	F(W_*) \in \Popt(W_*)=\sP_1.
\end{equation*}
For $\mu\in(0,1)$, we further have 
\begin{equation*}
	F(W_{1,\mu})=p_*^{(1)}
\end{equation*}
since $\Popt(W_{1,\mu})=\{p_*^{(1)}\}$ for $\mu\in(0,1)$. 

For $\mu\in(0,1)$ we also have 
\begin{equation*}
	F(W_{2,\mu})=p_*^{(2)}
\end{equation*}
since $\Popt(W_{2,\mu})=\{p_*^{(2)}\}$ for $\mu\in(0,1)$. We have $p_*^{(1)}\in\sP_1$, $p_*^{(2)}\in\sP_1$, and $\|p_*^{(1)}-p_*^{(2)}\|=1$. With this, we obtain
\begin{align*}
	1 &= \big\|p_*^{(1)} - p_*^{(2)}\big\|_{\ell_1} \\
	&= \big\|p_*^{(1)} - F(W_*) + F(W_*) - p_*^{(2)}\big\|_{\ell_1} \\
	&\leq \big\|p_*^{(1)} - F(W_*)\big\|_{\ell_1} + \big\|F(W_*) - p_*^{(2)}\big\|_{\ell_1} \\
	&\leq 2\max\Big\{ \big\|p_*^{(1)}-F(W_*)\big\|_{\ell_1}, \big\|p_*^{(2)}-F(W_*)\big\|_{\ell_1} \Big\}
\end{align*}
so that
\begin{equation*}
	\max\Big\{ \big\|p_*^{(1)}-F(W_*)\big\|_{\ell_1}, \big\|p_*^{(2)}-F(W_*)\big\|_{\ell_1} \Big\}\geq\frac{1}{2}.
\end{equation*}

Let $\sA\subset\N$ be a recursively enumerable set that is not recursive, cf. Section \ref{sec:computability}. Let $g:\N\rightarrow\sA$ be a computable function where for each $m\in\sA$ there exists a $k$ with $g(k)=m$ and $g(k_1)\neq g(k_2)$ for $k_1\neq k_2$. 

Let $\fT_\sA$ be a Turing machine that accepts exactly the set $\sA$, i.e., $\fT_\sA$ stops for input $k\in\N$ if and only if $k\in\sA$. Otherwise, $\fT_\sA$ runs forever and does not stop. For $k\in\N$ and $n\in\N$, we define the function
\begin{align*}
	q(k,n)\!=\!\! \begin{cases}
		\!2^{s+2}	&\!\!\!\!\text{if } \fT_\sA \!\text{ stops for input } k \text{ after } s\leq n \text{ steps} \\
		\!2^{n+2}	&\!\!\!\!\text{if } \fT_\sA \!\text{ does not stop for input } k \text{ after } n \text{ steps}.
	\end{cases}
\end{align*}
Note that $q:\N\times\N\rightarrow\N$ is a computable function.

Let $k,n\in\N$ be arbitrary. If $k$ is odd, i.e., $k\in\sO$ with $\sO\subset\N$ the set of all odd numbers, then we have $k=2l-1$, $l\geq1$, $l\in\N$, and we consider the channel $W_{k,n}\coloneqq W_{1,\frac{1}{q(l,n)}}$. If $k$ is even, i.e., $k\in\sE$ with $\sE\subset\N$ the set of all even numbers, then we have $k=2l$, $l\geq1$, $l\in\N$, and we consider $W_{k,n}\coloneqq W_{2,\frac{1}{q(l,n)}}$. Note that in both cases, $l$ is a function of $k$. With this, $(W_{k,n})_{k\in\N,n\in\N}$ is a computable double sequence. 

Now, we define the following sequence $(W_k^*)_{k\in\N}$. We will later show in the proof that $(W_k^*)_{k\in\N}$ is even a computable sequence of computable channels. For $k\in\N$, $k$ is either odd or even:
\begin{enumerate}
	\item $k\in\sO$ odd, i.e., $k=2l-1$, $l\geq1$, $l\in\N$. If $l\in\sA$, then we set $W_k^*\coloneqq W_{1,\frac{1}{2^{s+2}}}$ with $\fT_\sA$ has stopped for input $l$ after $s$ steps. If $l\notin\sA$, then we set $W_k^*\coloneqq W_*$.
	\item $k\in\sE$ even, i.e., $k=2l$, $l\geq1$, $l\in\N$. If $l\in\sA$, then we set $W_k^*\coloneqq W_{2,\frac{1}{2^{s+2}}}$ with $\fT_\sA$ has stopped for input $l$ after $s$ steps. If $l\notin\sA$, then we set $W_k^*\coloneqq W_*$.
\end{enumerate}
Next, we show that the double sequence $(W_{k,n})_{k\in\N,n\in\N}$ converges effectively to the sequence $(W_k^*)_{k\in\N}$. This implies that $(W_k^*)_{k\in\N}$ is a computable sequence of computable channels. Further, we show that for all $k\in\N$ and $n\in\N$ we have
\begin{equation}
	\label{eq:D}
	D(W_k^*,W_{k,n}) < \frac{1}{2^n}
\end{equation}
so that $(W_{k,n})_{k\in\N,n\in\N}$ indeed converges effectively. 

Let $k\in\N$ be arbitrary. We first consider the case $k\in\sO$, i.e., $k=2l-1$, $l\geq1$, $l\in\N$. If $l\notin\sA$, we have $W_k^*=W_*$ so that
\begin{align*}
	D\big(W_k^*,W_{k,n}\big) &= D\big(W_*, W_{1,\frac{1}{2^{n+2}}}\big) = \frac{2}{2^{n+2}} < \frac{1}{2^n}
\end{align*}
which already shows \eqref{eq:D} for this case. In the other case, if $l\in\sA$, we have $W_k^*=W_{1,\frac{1}{2^{s+2}}}$, where $s$ is the actual number of steps after which the Turing machine $\fT_\sA$ stopped for input $l$. Now, let $n\in\N$ be arbitrary. For $n\geq s$ we have
\begin{equation*}
	W_{k,n} = W_{2l-1,n} = W_{1,\frac{1}{2^{s+2}}} = W_k^*
\end{equation*}
so that
\begin{equation*}
	D(W_k^*,W_{k,n}) = 0.
\end{equation*}
For $n< s$ we have $W_{k,n}=W_{1,\frac{1}{2^{n+2}}}$ so that
\begin{align*}
	&D(W_k^*,W_{k,n}) = D(W_{1,\frac{1}{2^{s+2}}},W_{1,\frac{1}{2^{n+2}}}) \\
	&\qquad= \Big|\Big(1-\frac{1}{2^{s+2}}\Big)-\Big(1-\frac{1}{2^{n+2}}\Big)\Big| + \Big|\frac{1}{2^{s+2}}-\frac{1}{2^{n+2}}\Big| \\
	&\qquad= 2\Big|\frac{1}{2^{n+2}}-\frac{1}{2^{s+2}}\Big| < 2\frac{1}{2^{n+2}} < \frac{1}{2^n}
\end{align*}
which shows \eqref{eq:D} for this case as well. 

The proof for even numbers $k\in\sE$ follows as above for odd numbers $k\in\sO$ and is omitted for brevity. As a consequence, $(W_k^*)_{k\in\N}$ is a computable sequence of computable channels. If the function $F$ is Banach-Mazur computable, then the sequence $(F(W_k^*))_{k\in\N}$ must be a computable sequence of computable input distributions in $\sP_c(\sX)$. 

We consider the computable sequence
\begin{equation}
	\label{eq:F}
	\big(F(W_k^*)-F(W_*)\big)_{k\in\N}
\end{equation}
and the following Turing machine: For $l\in\N$ we start two Turing machines in parallel.

The first Turing machine $\fT_1$ is given by $\fT_1=\fT_\sA$, i.e., for input $l$ it runs the algorithm for $\fT_\sA$ step by step.

The second Turing machine is given as follows. We compute $n=2l-1$ and also $F(W_{2l-1}^*)-F(W_*)$ which is possible since \eqref{eq:F} is a computable sequence. We compute $\|F(W_{2l-1}^*)-F(W_*)\|_{\ell_1}$. In parallel, we further compute $n=2l$ and also $F(W_{2l}^*)-F(W_*)$ and $\|F(W_{2l}^*)-F(W_*)\|_{\ell_1}$. We now compute
\begin{align*}
	r_l = \max\big\{\|F(W_{2l-1}^*)\!-\!F(W_*)\|_{\ell_1},\|F(W_{2l}^*)\!-\!F(W_*)\|_{\ell_1}\big\}.
\end{align*}
We now use the Turing machine $\fT_{<\frac{1}{4}}$ from \cite[page 14]{PourElRichards-2017-ComputabilityAnalysisPhysics} and test if $r_l<\frac{1}{4}$ is true. Our second Turing machine $\fT_2$ stops if and only if the Turing machine $\fT_{<\frac{1}{4}}$ stops for input $r_l$. 

We start both Turing machines in parallel in such a way that the computing steps are synchronous. Whenever the first Turing machine stops, we set $l\in\sA$. Otherwise, if the second Turing machine stops, we set $l\notin\sA$. The first Turing machine stops if and only if $l\in\sA$. The second Turing machine stops if and only if $r_l<\frac{1}{4}$. As for $l\in\sA$ we have $r_l\geq\frac{1}{2}$ and for $l\notin\sA$ we have $r_l=0$, the second Turing machine stops if and only if $l\notin\sA$. 

With this, we have obtained a Turing machine $\fT_*$ that always decides for $l\in\N$ whether $l\in\sA$ or $l\notin\sA$. This means that $\sA$ must be a recursive set which is a contradiction to our initial assumption. Thus, the function $F$ is not Banach-Mazur computable which proves the desired result for the case $|\sX|=3$ and $|\sY|=2$.

\medskip

Finally, we outline how the proof extends to arbitrary $|\sX|\geq3$ and $|\sY|\geq2$. In this case, for the set $\CHcomp$ we consider the subset $\underline{\mathcal{CH}}_c(\sX;\sY)$ of all channels $W\in\CHcomp$ and choose an arbitrary channel $\underline{W}\in\CH_c(\sX_1;\sY_1)$ with $|\sX_1|=3$ and $|\sY_1|=2$. We set
\begin{align}
	\label{eq:construction1}
	W(y|x) = \begin{cases}
		\underline{W}(y|x)	&y\in\{1,2\} \and x\in\{1,2,3\} \\
		0	&y\in\{3,...,|\sY|\} \and x\in\{1,2,3\} 
	\end{cases}
\end{align}
as well as
\begin{equation}
	\label{eq:construction2}
	W(\cdot|x) = \underline{W}(\cdot|1) \quad x\in\{4,...,|\sX|\}.
\end{equation}

As above, we assume that $F\in\Mopt$ is a Banach-Mazur computable function that computes an optimal input distribution for the set $\CHcomp$. Then we always have $F(W)\in\sP(\sX)$ for $W\in\CHcomp$. For $\underline{W}\in\CH_c(\sX_1;\sY_1)$ we can immediately compute an optimal input distribution $p_1^*\in\Popt(\underline{W})$ as follows. We take $W$ which is constructed as above in \eqref{eq:construction1}-\eqref{eq:construction2}. Let $W\in\CHcomp$ and consider $p(W)\coloneqq F(W)$. With
\begin{equation*}
	p(W) = \begin{pmatrix}
		p_1(W) \\ \vdots \\ p_{|\sX|}(W)
	\end{pmatrix}
\end{equation*}
we set
\begin{equation}
	\label{eq:I1}
	p_1^*(\underline{W}) \coloneqq p_1(W) + \sum_{x=4}^{|\sX|}p_x(W)
\end{equation}
and
\begin{align}
	p_2^*(\underline{W}) &\coloneqq p_2(W), \label{eq:I2}\\
	p_3^*(\underline{W}) &\coloneqq p_3(W). \label{eq:I3}
\end{align}
For $\underline{W}\in\CH(\sX_1;\sY_1)$ we consider the mapping
\begin{equation*}
	G(\underline{W}) = \begin{pmatrix}
		p_1^*(\underline{W}) \\ p_2^*(\underline{W}) \\ p_3^*(\underline{W})
	\end{pmatrix}
\end{equation*}
which is defined by \eqref{eq:I1}-\eqref{eq:I3}. The mapping $G$ is a composition of the following components: 1) it constructs from $\underline{W}$ the channel $W$ according to \eqref{eq:construction1}-\eqref{eq:construction2}; 2) it applies the function $F$ on $W$; and 3) it applies the operations \eqref{eq:I1}-\eqref{eq:I3} on $F$. The construction \eqref{eq:construction1}-\eqref{eq:construction2} and also the operations \eqref{eq:I1}-\eqref{eq:I3} are Borel-Turing computable. Since $F$ is further Banach-Mazur computable by assumption, the mapping $G$ must be Banach-Mazur computable as well. However, we have $p_*\in\Popt(W)$. This is a contradiction since for $|\sX_1|=3$ and $|\sY_1|=2$ all functions $G\in\sM_{\text{opt}}(\sX_1;\sY_1)$ can not be Banach-Mazur computable. This proves the general case and therewith completes the proof of Theorem \ref{the:banachmazur}.
\end{IEEEproof}

\addspace

By inspection of the proof of Theorem \ref{the:banachmazur}, we observe that we have shown a stronger statement than was initially required. More specifically, we even have shown the following result:
	
\begin{theorem}
	\label{the:stronger}
	Let $\sX$ and $\sY$ be arbitrary finite alphabets. Then there exists a computable sequence $(W_n)_{n\in\N}$ of Borel-Turing computable channels, where every channel $W_n$, $n\in\N$ has only rational entries, such that for all $G\in\Mopt$ it always holds that $(G(W_n))_{n\in\N}$ is not a Borel-Turing computable sequence.
\end{theorem}
	
Some remarks are in order. We actually have a universal non-Banach-Mazur computability here. To show that a specific function $G\in\Mopt$ is not Banach-Mazur computable, we have to show that for this function $G$ there is a computable sequence $(W_n)_{n\in\N}$ such that $(G(W_n))_{n\in\N}$ is not a computable sequence of computable input distributions in $\sP(\sX)$. In general, the sequence $(W_n)_{n\in\N}$ depends on the function $G$. From a practical point of view, it could be the case that for an arbitrary given computable sequence $(W_n)_{n\in\N}$ of computable channels (which contain all practically relevant channels for certain application) there is a $G$ such that $G((W_n))_{n\in\N}$ becomes a computable sequence of output distributions $(p(W_n))_{n\in\N}$. However, we have shown that this possibility for the optimization of the mutual information is not possible, since Theorem \ref{the:stronger} excludes such a behavior since the sequence $(W_n)_{n\in\N}$ in Theorem \ref{the:stronger} is a universal sequence such that for all $G\in\Mopt$, the sequence $(G(W_n))_{n\in\N}$ is not a computable sequence. 

As discussed in the previous Section \ref{sec:system_problem}, we must not necessarily have $G(W)\in\sP_c(\sX)$ for $W\in\CHcomp$. In particular, already on the interval $[0,1]$ there are computable continuous functions $f$ such that for all optimizers $x_*\in[0,1]$ we have $x_*\notin\R_c$. However, we know that such a behavior cannot occur for the mutual information since for $|\sX|\geq2$ and $W\in\CHcomp$ arbitrary we have the following reasoning: $\Popt(W)$ is a non-empty set and if there is only one element in $\Popt(W)$, i.e., $|\Popt(W)|=1$, then we know from \cite[Sec. 0.6]{PourElRichards-2017-ComputabilityAnalysisPhysics} that for the optimal input distribution $p_*\in\sP_c(\sX)$ is satisfied. On the other hand, if $\Popt(W)$ contains more than one element, i.e., $|\Popt(W)|\geq2$, then $\Popt(W)$ is a convex set, i.e., for $p^{(1)},p^{(2)}\in\Popt(W)$ with $p^{(1)}\neq p^{(2)}$ we also have $p_\lambda\coloneqq (1-\lambda)p^{(1)}+\lambda p^{(2)}\in\Popt(W)$ for $\lambda\in[0,1]$.
	
Now let $i(1)$ be an arbitrary index of $\sX$ with $p^{(1)}(i(1))\neq p^{(2)}(i(1))$. For $p_\lambda(i(1))= (1-\lambda)p^{(1)}(i(1)) + \lambda p^{(2)}(i(1))$ we always have
\begin{equation*}
	p_\lambda(i(1)) \in [\underline{a}_1,\overline{a}^1]
\end{equation*}
with $\underline{a}_1<\overline{a}^1$ and
\begin{align*}
	\underline{a}_1 &= \min\big\{p^{(1)}(i(1)),p^{(2)}(i(1))\big\}, \\
	\overline{a}^1 &= \max\big\{p^{(1)}(i(1)),p^{(2)}(i(1))\big\}.
\end{align*}
This implies the existence of a $\hat{\lambda}\in(0,1)$ with $p_{\hat{\lambda}}(i(1))\in\R_c$.
	
Next, we consider $\sX_i=\sX\backslash\{i(1)\}$ and $\sM_1 = \{p\in\sP(\sX): p(i(1))=p_{\hat{\lambda}}(i(1))\}$ and study the relation
\begin{equation*}
	\max_{p\in\sM_1} I(p,W) = \max_{p\in\sP(\sX)} I(p,W).
\end{equation*}
Now, let $\Popt(W,\sM_1)$ be the set of all $p_*\sM_1$ with
\begin{equation*}
	I(p_*,W) = \max_{p\in\sM_1} I(p,W).
\end{equation*}
If $\Popt(W,\sM_1)$ consists of only one element, then we must have $p_*\in\sP_c(\sX)$. If $\Popt(W,\sM_1)$ consists of more than one element, then the set $\Popt(W,\sM_1)$ must be convex. In this case, we find another index $i(2)$ with $i(2)\neq i(1)$ such that for $i(2)$ there is a $\hat{p}_*\in\Popt(w,\sM_1)$ with $\hat{p}_*\in\Popt(W,\sM_1)$ and $\hat{p}_*(i(2))\in\R_c$. This procedure can be continued iteratively such that the index set is reduced by one element in each iteration. After finitely many steps, we obtain a $\tilde{p}\in\Popt(W)$ with $\tilde{p}\in\sP_c(\sX)$. 
	
It is clear that this procedure allows us to show the existence of such a $\tilde{p}$ only, but that $\tilde{p}$ cannot be constructed algorithmically. It is interesting to note that this allows us to show the existence of such a function $G$ with $G(W)=\Popt(W)$ for all $W\in\CHcomp$ such that we always have $G(W)\in\sP_c(\sX)$ for $W\in\CHcomp$ is true. Accordingly, for $W\in\CHcomp$ it is impossible that $\Popt(W)$ contains only non-Borel-Turing computable input distributions.
	
Finally, we want to emphasize that it has recently been shown for other information theoretic problems arising in prediction theory that for simple spectral power densities the corresponding prediction filters and Wiener filter, accordingly, for Borel-Turing computable frequences have non-Borel-Turing computable values \cite{BochePohl}. In contrast to this, such a behavior cannot occur for the optimal input distribution in our case.

% ======================================================================================================
\subsection{Discussion}

Some discussion is in order.

\begin{remark}
	This shows that such a Turing machine cannot exist providing a negative answer to Question 1 above. As a consequence, this means also that a function $F$ as in \eqref{eq:system_f} cannot exist for which $F(W)$ can ``easily'' be computed for $W$. In particular, this excludes the possibility of finding a function $F$ that provides a ``closed form solution'', since this would be then Turing computable and therewith algorithmically constructable, cf. also \cite{Chow-1999-ClosedFormNumber,BorweinCrandall-2013-ClosedForms}. 
\end{remark}

\begin{remark}
	It is of interest to discuss the Blahut-Arimoto algorithm taking the result in Theorem \ref{the:banachmazur} into account. This algorithm computes for each channel $W\in\CHcomp$ a sequence $(p_n(W))_{n\in\N}$ of input distributions such that all convergent subsequences always converge to a corresponding optimizer $p_*(W)\in\Popt(W)$. The second crucial ingredient of the representation of $p_*(W)$ is a stopping criterion for the computation of the sequence $(p_n)_{n\in\N}$ for a given approximation error $\frac{1}{2^N}$. Such a stopping criterion is not provided by the Blahut-Arimoto algorithm. This was already criticized by Csisz\'ar in \cite{Csiszar-1974-ComputationRateDistortion}. Our Theorem~\ref{the:banachmazur} shows now that such a computable stopping criterion as a function of the representation of the channel cannot exist.
\end{remark}

\begin{remark}
	Theorem \ref{the:banachmazur} further shows that the convergence behavior of the Blahut-Arimoto algorithm cannot be improved by optimizing the starting point of the algorithm, i.e., by choosing the starting point in the form of a Turing computable pre-processing such that it results in a computable stopping criterion for the algorithm. As a consequence, there is no Turing computable function $G_0:\CHcomp\rightarrow\Popt(W)$ with $p_0(W)=G_0(W)$ such that a Turing computable stopping criterion would exist for the Blahut-Arimoto algorithm with $p_0(W)$ as initialization. 
\end{remark}

The statement on the impossibility of the algorithmic solvability is closely connected to the underlying hardware platform (Turing machine) and therewith, equivalently, to the admissible programming languages (Turing complete) and also the admissible signal processing operations. Note that for other computing platforms (such as neuromorphic or quantum computing platforms) this statement need not be the case. However, whenever simulations are done in the broad area of information theory, communication theory, or signal processing, these are done on digital hardware platforms for which Turing machines provide the underlying computing framework. 

\begin{remark}
	It is helpful and very interesting to gain further intuition and insight into the non-computability by Turing machines and other potential computing platforms. For example, it has been a long-standing open problem to describe the roots of polynomials by radicals as a function of the coefficients of the polynomial. To this end, Galois showed this is not possible in general for polynomials of the order 5 or higher \cite{Herstein-1975-TopicsInAlgebra}. This means that the roots of polynomials of order 5 or higher cannot be expressed as a ``closed form solution'' by radicals; see \cite{Herstein-1975-TopicsInAlgebra} and further discussions in \cite{Chow-1999-ClosedFormNumber,BorweinCrandall-2013-ClosedForms}. On the other hand, from the complex analysis there are algorithms known that are able to approximate these roots. This shows that the ``computing theory of radicals'' is not sufficient for the computation of the roots of polynomials of order 5 or higher, but other techniques from complex analysis enable the approximation thereof. 
\end{remark}

Next, we want to further discuss the implications of Theorem \ref{the:banachmazur} and the problem of the computation of $C(W)$, $W\in\CHcomp$. For this purpose, let $W\in\CHcomp$ be fixed and we consider the function $f_W(p)\coloneqq I(p,W)$. The function $f_W$ is concave and further a computable function in $p$ since $W\in\CHcomp$. Therefore, from \cite[Section 0.6]{PourElRichards-2017-ComputabilityAnalysisPhysics} follows that the condition $C(W)\in\R_c$ is satisfied, cf. also \eqref{eq:B}. For $p_*\in\Popt(W)$ arbitrary we have $f_W(p_*)=C(W)$, but $p_*$ must not necessarily be a computable input distribution, i.e., $p_*\in\sP_c(\sX)$ is not necessarily satisfied. In fact, in \cite{Specker-1959-Maximum} it has been shown that already on the interval $[0,1]$ it is possible to construct continuous computable functions $g$ such that for every point $x_*\in[0,1]$ with $\max_{x\in[0,1]}g(x)=g(x_*)$ it holds that $x_*\notin\R_c$, i.e., there is no Borel-Turing computable maximizer, although we have $\max_{x\in[0,1]}g(x)\in\R_c$. 

For the mutual information, i.e., for our function $f_W$, such a behavior cannot exist. This means there always exists an optimal $p_*\in\Popt(W)$ such that $p_*\in\sP_c(\sX)$ is satisfied. This implies the following: For an optimal $p_*\in\sP_c(\sX)\cap\Popt(W)$ there exists by definition an algorithm (or Turing machine) that approximates the optimal $p_*$ by rational probability distributions with arbitrarily small approximation error. In particular, with this we can find a function $G_*$ such that for every $W\in\CHcomp$ it always holds $G_*(W)\in\sP_c(\sX)\cap\Popt(W)$. This means the function $G_*$ is mathematically well defined and gives always a computable optimal input distribution, i.e., the output of $G_*$ is in $\sP_c(\sX)$. But, at the same time, the function $G_*$ itself is not Borel-Turing computable, i.e., the function $W\rightarrow p_*(W)\in\sP_c(\sX)\cap\Popt(W)$ does not depend on $W$ in a Borel-Turing computable way. In particular, for every $W\in\CHcomp$ there exists a $p_*(W)\in\sP_c(\sX)\cap\Popt(W)$, but this $p_*(W)$ cannot be computed algorithmically. 

\begin{remark}
	For $W\in\CH(\sX;\sY)$ the set $\Popt$ is always convex. For practical applications, it would be interesting to know the extreme points of this set. Then it is of further interest to understand if for $W\in\CHcomp$ the extreme points of the set $\Popt$ are also in $\sP_c(\sX)$, i.e., a computable input distribution. This question remains open and for general continuous computable concave functions it could be the case that no extreme point of $\Popt$ is computable.
\end{remark}

% ========================================================================================
\section{Approximability of an Optimal Input Distribution}
\label{sec:system_approx}

\begin{figure}
	\centering
	\scalebox{1}{\includegraphics{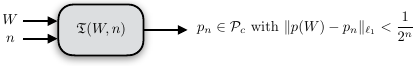}}
	\caption{The Turing machine $\fT$ obtains a description of the channel $W$ and the block length $n$ as inputs and outputs a description of an input distribution $p(W)$ that satisfies the tolerated approximation error $\frac{1}{2^n}$. Here, an algorithmic dependency on the channel $W$ is given as $W$ is provided as an input to $\fT$.}
	\label{fig:2}
\end{figure}

\begin{figure}
	\centering
	\scalebox{1}{\includegraphics{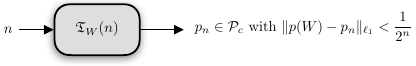}}
	\caption{The Turing machine $\fT$ obtains the block length $n$ as input and outputs a description of an input distribution $p(W)$ that satisfies the tolerated approximation error $\frac{1}{2^n}$. Here, the Turing machine $\fT_W$ depends on $W$ in the sense that for every $W\in\CHcomp$ there exists a Turing machine with the desired properties.}
	\label{fig:3}
\end{figure}

Above we have shown that it is impossible to algorithmically construct optimal, i.e., capacity-achieving, input distributions. Consequently, we are now interested in understanding whether or not it is at least possible to algorithmically approximate such distributions. This is visualized in Fig.~\ref{fig:2} for the case where the Turing machine would obtain both the channel and the block length as inputs and in Fig.~\ref{fig:3} for the case where the channel is known beforehand and only the block length is given as an input to the Turing machine.

We have seen that all functions $F:\CHcomp\rightarrow\sP(\sX)$ with $F(W)\in\Popt(W)$ for all $W\in\CHcomp$ are not Banach-Mazur computable and therewith also not Borel-Turing computable. The question is now whether or not we can instead solve this problem approximately, i.e., does there exist a computable sequence of Borel-Turing computable functions $F_n$, $n\in\N$,  with $F_n:\CHcomp\rightarrow\sP_c(\sX)$, $n\in\N$, such that for all $W\in\CHcomp$ for a suitable function $F$ with $F(W)\in\Popt(W)$ for all $W\in\CHcomp$ we always have
\begin{equation*}
	\big\|F(W)-F_n(W)\big\|_{\ell_1} < \frac{1}{2^n}.
\end{equation*}
This is equivalent to the question of whether or not there exists a Turing machine $\fT$ that takes an arbitrary representation of $W\in\CHcomp$ and $n\in\N$ as inputs and then computes for $W$ and $n$ a representation for $p_n(W)\in\sP(\sX)$ such that 
\begin{equation}
	\label{eq:approx_error}
	\big\|F(W)-p_n(W)\big\|_{\ell_1} < \frac{1}{2^n}.
\end{equation}
And this is equivalent to the question of whether or not it is possible to find a Turing machine $\fT$ with the following properties: $\fT$ takes the channel and natural numbers as inputs and computes a description of an input distribution. This input distribution must satisfy the following: for all $W\in\CHcomp$ and all $n\in\N$ the Turing machine must compute for every description for $W$ a description of $p_n(W)$ such that for a suitable $p_*(W)\in\Popt(W)$ it always holds that
\begin{equation*}
	\big\|p_*(W)-p_n(W)\big\|_{\ell_1} < \frac{1}{2^n}.
\end{equation*}
The input $n$ of this Turing machine $\fT$ could enable the algorithmic approximation of the optimal input distribution.

A negative answer can be immediately given to this question based on the results obtained above, since a function $F$ must be Borel-Turing computable, see also \cite{BocheSchaeferPoor-2020-CIS-NonComputabilityFSC}. We can formalize the following question.
\addspace

\begin{tcolorbox}[colback=white,boxrule=0.125ex]
	{\bf Question 2:} 
	Let $\sX$ and $\sY$ be finite input and output alphabets with $|\sX|\geq3$ and $|\sY|\geq2$. Is it possible to approximate a function $F\in\Mopt$ by computable functions? Is there a function $F\in\Mopt$ and a computable function $F_1$ such that
	\begin{equation*}
		\sup_{W\in\CHcomp}\big\|F(W)-F_1(W)\big\|_{\ell_1} < \frac{1}{2}\;?
	\end{equation*}
\end{tcolorbox}

\begin{remark}
	\label{rem:question2}
	With this question we ask whether or not the previous condition \eqref{eq:approx_error} as the supremum can be satisfied for the trivial case $n=1$.
\end{remark}

\begin{theorem}
	\label{the:approx}
	Let $\sX$ and $\sY$ be arbitrary but finite alphabets with $|\sX|\geq3$ and $|\sY|\geq2$. Let $F\in\Mopt$ be an arbitrary function and let $F_1$ be another arbitrary function with
	\begin{equation*}
		\sup_{W\in\CHcomp}\big\|F(W)-F_1(W)\big\|_{\ell_1} = \alpha < \frac{1}{2}.
	\end{equation*}
	Then $F_1$ is not Banach-Mazur computable. 
\end{theorem}
\begin{IEEEproof}
	We prove the result by contradiction. Therefore, we assume that there exists a function $F\in\Mopt$ such that there is a function $F_1$ with 
	\begin{equation*}
		\sup_{W\in\CHcomp}\big|F(W)-F_1(W)\big| = \beta < 1
	\end{equation*}
	that is Banach-Mazur computable. Then, there exists a computable real number $\alpha$ with $\beta\leq\alpha<1$. 
	
	We now consider the computable sequence $(W_n^*)_{n\in\N}$ as used in the proof of Theorem \ref{the:banachmazur}. For $l\in\N$, let
	\begin{equation*}
		\big\| F_1(W_{2l}^*) - F(W_{2l}^*) \big\|_{\ell_1} \leq \alpha
	\end{equation*} 
	and
	\begin{equation*}
		\big\| F_1(W_{2l-1}^*) - F(W_{2l-1}^*) \big\|_{\ell_1} \leq \alpha
	\end{equation*} 
	be satisfied. Then, we also have for $l\in\sA$ the following:
	\begin{align*}
		1 &= \big\| F(W_{2l}^*) - F(W_{2l-1}^*) \big\|_{\ell_1} \\
		&= \big\| F(W_{2l}^*) - F_1(W_{2l}^*) + F_1(W_{2l}^*) - F_1(W_{2l-1}^*)\\
		&\qquad + F_1(W_{2l-1}^*)- F(W_{2l-1}^*) \big\|_{\ell_1} \\
		&\leq \big\| F(W_{2l}^*) - F_1(W_{2l}^*)\big\|_{\ell_1} + \big\|F_1(W_{2l}^*) - F_1(W_{2l-1}^*)\big\|_{\ell_1}\\
		&\qquad + \big\|F_1(W_{2l-1}^*)- F(W_{2l-1}^*) \big\|_{\ell_1} \\
		&\leq 2\alpha + \big\|F_1(W_{2l}^*)- F_1(W_{2l-1}^*) \big\|_{\ell_1}.
	\end{align*}
	Therefore, it holds that
	\begin{equation*}
		\big\|F_1(W_{2l}^*)- F_1(W_{2l-1}^*) \big\|_{\ell_1} \geq 1-2\alpha = c_1>0
	\end{equation*}
	which implies that
	\begin{align*}
		c_1 &= \big\|F_1(W_{2l}^*) - F_1(W_*) + F_1(W_*) - F_1(W_{2l-1}^*) \big\|_{\ell_1} \\
		&\leq \big\|F_1(W_{2l}^*) - F_1(W_*)\big\|_{\ell_1} + \big\|F_1(W_*) - F_1(W_{2l-1}^*) \big\|_{\ell_1} \\
		&\leq 2\max\big\{\big\|F_1(W_{2l}^*) \!-\! F_1(W_*)\big\|_{\ell_1} , \\
		&\qquad\quad\big\|F_1(W_*) \!-\! F_1(W_{2l-1}^*)\big\|_{\ell_1}\big\} \\
		&\eqqcolon 2r_l^*.
	\end{align*}
	We conclude that
	\begin{align}
		\label{eq:rl}
		r_l^* \geq \frac{c_1}{2}>0.
	\end{align}
	
	For $l\in\N$ and $l\notin\sA$, 
	\begin{equation*}
		F_1(W_{2l}^*) = F_1(W_*)
	\end{equation*}
	and 
	\begin{equation*}
		F_1(W_{2l-1}^*) = F_1(W_*)
	\end{equation*}
	are satisfied. Accordingly, we can use the same Turing machine $\fT_{<\frac{1}{4}}$  as in the proof of Theorem \ref{the:banachmazur} for the input $r_l^*$ in \eqref{eq:rl}. The Turing machine $\fT_{<\frac{1}{4}}(r_l^*)$ stops if and only if $l\notin\sA$. Thus, we can construct a Turing machine as in the proof of Theorem \ref{the:banachmazur} that decides for every $l\in\N$ whether $l\in\sA$ or $l\notin\sA$. This is, again, a contradiction to the initial assumption completing the proof of Theorem \ref{the:approx}.
\end{IEEEproof}
\addspace

From this we immediately obtain the following result. 

\begin{corollary}
	\label{cor:approx}
	Let $F\in\Mopt$ be an arbitrary function. For $\alpha<\frac{1}{2}$ arbitrary, there exists no Turing machine $\fT_*$ such that for all $W\in\CHcomp$,
	\begin{equation*}
		\big\| F(W) - \fT_*(W) \big\|_{\ell_1} \leq \alpha
	\end{equation*} 
	is true.
\end{corollary}
\begin{IEEEproof}
	If such a function $F\in\Mopt$ would exist for which we can find a Turing machine $\fT_*$ with $\hat{\alpha}<\frac{1}{2}$, then $F_1(W)=\fT_*(W)$, $W\in\CHcomp$, would be Banach-Mazur computable. 
\end{IEEEproof}
\addspace

As a consequence, we can further conclude the following.

\begin{corollary}
	\label{cor:approx2}
	The approximation problem stated in Question 2 is not solvable. 
\end{corollary}
\begin{IEEEproof}
	Already for $n=2$ this is not possible. 
\end{IEEEproof}
\addspace

Similarly as in Section \ref{sec:system_problem}, one can show that a choice of Borel-Turing computable starting points for iterative algorithms such as the Blahut-Arimoto algorithm, i.e., $p_0(W)=G(W)$, $W\in\CHcomp$, $G$ Borel-Turing computable function, does not improve the approximation behavior according to Question 2.

% ======================================================================================================
% ======================================================================================================
% ======================================================================================================
\section{Conclusion}
\label{sec:conclusion}

The channel capacity describes the maximum rate at which a source can be reliably transmitted. Capacity expressions are usually given by entropic quantities that are optimized over all possible input distributions. Evaluating such capacity expressions and finding corresponding optimal input distributions that maximize these capacity expressions is a common and important task in information and communication theory. Several algorithms including the Blahut-Arimoto algorithm have been proposed to algorithmically compute these quantities. In this work, we have shown that there exists no algorithm or Turing machine that takes a DMC as input and then computes an input distribution that maximizes the capacity. Although capacity-achieving input distributions have been found analytically for some specific DMCs, this does not immediately mean that capacity-achieving input distributions can be algorithmically computed by a Turing that takes a DMC of interest as input. We have further shown that it is not even possible to algorithmically approximate this distribution. These results have implications for the Blahut-Arimoto algorithm. In particular, as we have noted, there is no stopping criterion for the computation of the input distribution, and our results imply that such a computable stopping criterion cannot exist, providing a negative answer to the open question of whether one does.

Future communication systems such as the 6th generation (6G) of mobile communication networks are being developed for very critical applications such as autonomous driving or mobile robots, but also for health care \cite{FettweisBoche-2021-BITS-PersonalTactileInternet,Schwenteck-2023-NetLet-TactileInternet}. Due to the particular challenges of security and privacy of these applications, 6G must fulfill strict conditions on trustworthiness \cite{FettweisBoche-2022-ProcACM-6GTrust}, for which integrity is one of the essential conditions that needs to be satisfied. With the results of this work, it can be shown that the computation of optimal input distributions is never possible on digital computers under the condition of integrity. In addition to the technical requirement for integrity, algorithms must also fulfill legal requirements for many future applications. For example, algorithms for critical decision problems must already fulfill the legal requirement of algorithmic transparency\footnote{Algorithmic transparency requires all factors that determine the result of an algorithm to be visible to the legislator, operator, user, and other affected individuals.}. With the results of this work, it can further be shown that the computation of optimal input distributions on digital computers is never possible under the requirement of algorithmic transparency.

\section*{Acknowledgment}

Holger Boche thanks Rainer Moorfeld for discussions and hints on new computing technologies for future communication systems. He also thanks Gerhard Fettweis for introducing him to the large field of achieving trustworthiness in 6G.

% ======================================================================================================
% ======================================================================================================
% ======================================================================================================
\appendix

% ======================================================================================================
\subsection{Example of a Non-Computable Function}
\label{app:function}

Here, we show that for $x\in[0,1]\cap\R_c$ the function
\begin{equation*}
	f(t) = e^{-x}
\end{equation*}
is not exactly computable on Turing machines, but only approximable. 

By the remainder theorem of Lagrange, we get for $x\in[0,1]$:
\begin{equation*}
	f(x) = \sum_{l=0}^n \frac{(-1)^l}{l!}x^l + \frac{1}{(n+1)!}f^{(n+1)}(\vartheta_x)x^{n+1}
\end{equation*}
with $\vartheta_x\in[0,x]$ a suitable number. With this, we get
\begin{equation*}
	\Big|f(x) - \sum_{l=1}^n\frac{(-1)^l}{l!}x^l\Big| = \frac{1}{(n+1)!}e^{-\vartheta_x}x^{n+1} \leq \frac{1}{(n+1)!}
\end{equation*}
and
\begin{equation*}
	(n+1)! > 2^n, \qquad n\geq 2.
\end{equation*}
Assume now that we have a sequence $(r_n)_{n\in\N}$ of rational numbers with
\begin{equation*}
	|x-r_n| < \frac{1}{2^n}
\end{equation*}
so that
\begin{align*}
	&\Big|f(x)-\sum_{l=0}^n\frac{(-1)^l}{l!}(r_n)^l\Big| \\
	&\qquad= \Big|f(x)-f(r_n)+f(r_n)-\sum_{l=0}^n\frac{(-1)^l}{l!}(r_n)^l\Big| \\
	&\qquad\leq |f(x)-f(r_n)| + \Big|f(r_n)-\sum_{l=0}^n\frac{(-1)^l}{l!}(r_n)^l\Big| \\
	&\qquad< |f(x)-f(r_n)| + \frac{1}{2^n}, \qquad n\geq2.
\end{align*}
Now, the mean value theorem implies that
\begin{equation*}
	|f(x)-f(r_n)| = |f'(\xi_{x,n})|\!\cdot\!|x-r_n|
\end{equation*}
with $\xi_{x,n}\in[x-r_n,x+r_n]$ being a suitable number. This yields
\begin{equation*}
	|f(x)-f(r_n)| \leq |x-r_n| < \frac{1}{2^n}.
\end{equation*}
With $y_n := \sum_{l=0}^n\frac{(-1)^l}{l!}(r_n)^l$, we obtain
\begin{equation*}
	|f(x)-y_n| < \frac{1}{2^n}+\frac{1}{2^n} = \frac{1}{2^{n-1}},
\end{equation*}
i.e., the algorithm
\begin{equation*}
	(r_n)_{n\in\N} \rightarrow (y_n)_{n\in\N}
\end{equation*}
maps a representation of $x$ into a representation of $f(x)$. This algorithm converges effectively.

From this calculation we immediately see how the function $f$ can be approximated. Whenever $f$ needs to be approximated in such a way that the error satisfies $\frac{1}{2^n}$, we use the polynomial as given above and compute it accordingly. For this, it is important to find suitable sequences of polynomials. Note that the polynomials in this sequence needs to be computable as well as the sequence itself needs to be a computable sequence, since otherwise, we are not able to evaluate the approximation process algorithmically. Note that this does not mean that every sequence of approximations of $f$ is also a suitable sequence for our purpose.

% ======================================================================================================
\subsection{Binary Entropy and Transcendental Numbers}
\label{app:transcendental}

For the following, we need Hilbert's Seventh Problem which is restated next for completeness.
\addspace

\begin{tcolorbox}[colback=white,boxrule=0.125ex]
	{\bf Hilbert's Seventh Problem.}
	 Let $a\notin\{0,1\}$ be an algebraic number (i.e., a root of a non-zero polynomial with integer coefficients) and let $b$ be an irrational and algebraic number. Is $a^b$ always a transcendental number (i.e., not algebraic)?
\end{tcolorbox}

\begin{remark}
A positive answer to this question was then first given in 1934 by Gelfond \cite{Gelfond-1934-Hilbert} and subsequently refined in 1935 by Schneider \cite{Schneider-1935-Hilbert}. Later this was generalized by Baker for which he was awarded a Fields Medal in 1970, cf. \cite{Baker-1975-TranscendentalNumberTheory}.
\end{remark}

We further need the following observation.

\begin{lemma}
	\label{lem:fakt1}
	Let $n\in\N$ and $t\in\N$ be arbitrary. Then $n$ and $n^t$ are divisible by the same prime numbers.
\end{lemma}
\begin{IEEEproof}
	Let $n=\prod_{l=1}^rp_l$ be the unique prime factorization of $n$. Note that in factorization, certain prime factors may appear multiple times. Then, $n^t=\prod_{l=1}^r(p_l)^t$ is a prime factorization of $n^t$. As this factorization is unique, both $n$ and $n^t$ must have the same prime factors.
\end{IEEEproof}
\addspace

We now prove the following result. 

\begin{theorem}
	\label{the:app}
	Let $p\in\Q$ with $p\notin\{0,\frac{1}{2},1\}$. Then, $h_2(p)$ is a transcendental number.
\end{theorem}
\begin{IEEEproof}
	Let 
	\begin{equation*}
		h_2(p) = p\log \frac{1}{p} + (1-p)\log\frac{1}{1-p}
	\end{equation*}
	be the binary entropy, which can be equivalently be expressed as
	\begin{equation}
		\label{eq:app_2h2p}
		2^{h_2(p)} = \left(\frac{1}{p}\right)^p\left(\frac{1}{1-p}\right)^{1-p}.
	\end{equation}
	Let $p\in\Q$ with $p\in(0,1)$, $p\notin\{0,\frac{1}{2},1\}$ be arbitrary. Then, we can express $p$ as $p=\frac{n}{m}$, $n<m$, $n,m\in\N$, and assume without loss of generality that $n$ and $m$ are coprime. With this, we can write
	\begin{equation*}
		\left(\frac{1}{p}\right)^p = \left(\frac{m}{n}\right)^\frac{n}{m}
	\end{equation*}
	and conclude that the number $(\frac{1}{p})^p$ is a root of the polynomial $x^m-(\frac{m}{n})^n$ and therewith also of the polynomial $n^nx^m-m^n$. Thus, $(\frac{1}{p})^p$ is an algebraic number. Similarly, one can show that $(\frac{1}{1-p})^{1-p}$ is an algebraic number so that $2^{h_2(p)}$ as in \eqref{eq:app_2h2p}	is also an algebraic number. 
	
	Now, we can use the Gelfond-Schneider theorem, i.e., the solution to Hilbert's Seventh Problem, cf. for example \cite{Baker-1975-TranscendentalNumberTheory}. As $2^{h_2(p)}$ is an algebraic number, $h_2(p)$ must be either rational or transcendental. Since otherwise, if $h_2(p)$ would be algebraic and irrational, then $2^{h_2(p)}$ would be transcendental. 
	
	Next, we want to show by contradiction that $h_2(p)$ cannot be rational. Since $p\in(0,1)$, $p\neq\frac{1}{2}$, we have $h_2(p)\in(0,1)$. For this purpose, we assume that $h_2(p)$ is rational so that it can be expressed as $h_2(p)=\frac{u}{v}$ with $0<u<v$, $u,v\in\N$, and $u,v$ coprime without loss of generality. We further must have $v>1$. This would imply that
	\begin{align*}
		2^{\frac{u}{v}} &= \left(\frac{m}{n}\right)^{\frac{n}{m}}\left(\frac{1}{1-\frac{n}{m}}\right)^{1-\frac{n}{m}} \\
			&= \left(\frac{m}{n}\right)^{\frac{n}{m}}\left(\frac{m}{m-n}\right)^{\frac{m-n}{m}}
	\end{align*}
	so that
	\begin{equation*}
		2^{mu} = \left(\frac{m}{n}\right)^{nv}\left(\frac{m}{m-n}\right)^{(m-n)v}
	\end{equation*}
	or equivalently
	\begin{equation*}
		2^{mu}(n)^{nv}(m-n)^{(m-n)v}= (m)^{mv}.
	\end{equation*}
	Note that $m-n\geq1$ and further $nv\in\N$, $nv>1$, since $v>1$. 
	
	If $n=1$, then
	\begin{equation*}
		2^{mu}(m-1)^{(m-1)v}= (m)^{mv}.
	\end{equation*}
	Lemma \ref{lem:fakt1} and the uniqueness of the prime factorization would then imply that every prime factor of $m-1$ must be a prime factor $m$ as well. However, this is not possible. 
	
	If $n>1$, then Lemma \ref{lem:fakt1} implies that every prime factor of $n$ must also be a prime factor of $m$. However, this is not possible, since $n$ and $m$ are coprime. As a consequence, $h_2(p)$ cannot be a rational number. Finally, we conclude that $h_2(p)$ must be a transcendental number which completes the proof.
\end{IEEEproof}

% ========================================================================================
% ========================================================================================
% ========================================================================================
\balance


\begin{IEEEbiographynophoto}{Holger Boche} (Fellow, IEEE) received the Dipl.-Ing. degree in electrical engineering, Graduate degree in mathematics, and the Dr.-Ing. degree in electrical engineering from the Technische Universit\"at Dresden, Dresden, Germany, in 1990, 1992, and 1994, respectively.  From 1994 to 1997, he did postgraduate studies at the Friedrich-Schiller Universit\"at Jena, Jena, Germany. In 1998, he received the Dr. rer. nat. degree in pure mathematics from the Technische Universit\"at Berlin, Berlin, Germany. 
	
	In 1997, he joined the Fraunhofer Institute for Telecommunications, Heinrich-Hertz-Institute (HHI), Berlin, Germany. From 2002 to 2010, he was Full Professor in mobile communication networks with the Institute for Communications Systems, Technische Universitä\"at Berlin. In 2003, he became the Director of the Fraunhofer German-Sino Laboratory for Mobile Communications, Berlin, and in 2004, he became the Director of the Fraunhofer Institute for Telecommunications (HHI). He was a Visiting Professor with the ETH Zurich, Zurich, Switzerland, during 2004 and 2006 (Winter), and with KTH Stockholm, Stockholm, Sweden, in 2005 (Summer). He is currently Full Professor at the Institute of Theoretical Information Technology, Technische Universit\"at M\"unchen, Munich, Germany, which he joined in October 2010. Among his publications is the book \emph{Information Theoretic Security and Privacy of Information Systems} (Cambridge University Press, 2017).
	
	Since 2014, Prof. Boche has been a member and Honorary Fellow of the TUM Institute for Advanced Study, Munich, Germany, and since 2018, a Founding Director of the Center for Quantum Engineering, Technische Universit\"at M\"unchen. Since 2021, he has been leading jointly with Frank Fitzek the research hub 6G-life. He is a member of IEEE Signal Processing Society SPCOM and SPTM Technical Committees. He was elected member of the German Academy of Sciences (Leopoldina) in 2008 and to the Berlin Brandenburg Academy of Sciences and Humanities in 2009. He is a recipient of the Research Award ``Technische Kommunikation'' from the Alcatel SEL Foundation in October 2003, the ``Innovation Award'' from the Vodafone Foundation in June 2006, and the Gottfried Wilhelm Leibniz Prize from the Deutsche Forschungsgemeinschaft (German Research Foundation) in 2008. He was a co-recipient of the 2006 IEEE Signal Processing Society Best Paper Award and a recipient of the 2007 IEEE Signal Processing Society Best Paper Award. He was General Chair of the Symposium on Information Theoretic Approaches to Security and Privacy at IEEE GlobalSIP 2016.
\end{IEEEbiographynophoto}

\begin{IEEEbiographynophoto}{Rafael F. Schaefer} (Senior Member, IEEE) is a Professor and head of the Chair of Information Theory and Machine Learning at Technische Universit\"at Dresden. He received the Dipl.-Ing. degree in electrical engineering and computer science from the Technische Universit\"at Berlin, Germany, in 2007, and the Dr.-Ing. degree in electrical engineering from the Technische Universit\"at M\"unchen, Germany, in 2012. From 2013 to 2015, he was a Post-Doctoral Research Fellow with Princeton University. From 2015 to 2020, he was an Assistant Professor with the Technische Universit\"at Berlin, Germany, and from 2021 to 2022 a Professor with the Universit\"at Siegen, Germany. Among his publications is the book \emph{Information Theoretic Security and Privacy of Information Systems} (Cambridge University Press, 2017). He was a recipient of the VDE Johann-Philipp-Reis Award in 2013. He received the best paper award of the German Information Technology Society (ITG-Preis) in 2016. He is currently an Associate Editor of the \textsc{IEEE TRANSACTIONS ON INFORMATION FORENSICS AND SECURITY} and of the \textsc{IEEE TRANSACTIONS ON COMMUNICATIONS}. 
\end{IEEEbiographynophoto}

\begin{IEEEbiographynophoto}{H. Vincent Poor} (Life Fellow, IEEE) received the Ph.D. degree in EECS from Princeton University in 1977.  From 1977 until 1990, he was on the faculty of the University of Illinois at Urbana-Champaign. Since 1990 he has been on the faculty at Princeton, where he is currently the Michael Henry Strater University Professor. During 2006 to 2016, he served as the dean of Princeton’s School of Engineering and Applied Science. He has also held visiting appointments at several other universities, including most recently at Berkeley and Cambridge. His research interests are in the areas of information theory, machine learning and network science, and their applications in wireless networks, energy systems and related fields. Among his publications in these areas is the recent book \emph{Machine Learning and Wireless Communications}  (Cambridge University Press, 2022). Dr. Poor is a member of the National Academy of Engineering and the National Academy of Sciences and is a foreign member of the Chinese Academy of Sciences, the Royal Society, and other national and international academies. He received the IEEE Alexander Graham Bell Medal in 2017.
\end{IEEEbiographynophoto}
\end{document}